\documentclass[a4paper,preprint, prl,showpacs]{revtex4}

\usepackage{amsmath,amssymb} 
\usepackage{makeidx}
\usepackage{amsfonts}
\usepackage[ansinew]{inputenc}
\usepackage[usenames,dvipsnames]{pstricks}
\usepackage{epsfig,graphicx}

\DeclareMathOperator*{\sgn}{sgn}
	
\newcommand{\vecSpin}{\mbox{\boldmath$S$}} 
\newcommand{\vecSpinhat}{\hat\vecSpin}
\newcommand{\vecPsi}{\mbox{\boldmath$\psi$}}
\newcommand{\vecPsihat}{\hat\vecPsi}

\newcommand{\Spin}{\mbox{$S$}}
\newcommand{\vecProb}{\mbox{\boldmath$P$}} 
\newcommand{\vecProbhat}{\hat\vecProb}
\newcommand{\Prob}{\mbox{$P$}}
\newcommand{\Probhat}{\hat{\Prob}}
\newcommand{\M}{\mbox{$M$}}
\newcommand{\vecOmega}{\mbox{\boldmath $\Omega$}} 
\newcommand{\vecOmegahat}{\hat\vecOmega}
\newcommand{\Psihat}{\hat\psi}
\newcommand{\betahat}{\hat\beta}
\newcommand{\nullvec}{\mbox{\boldmath$0$}}
\newcommand{\rme}{\mathrm{e}}
\newcommand{\rmi}{\mathrm{i}}
\newcommand{\rmd}{\mathrm{d}}

\makeindex

\begin{document}

\title{Computing with Noise - Phase Transitions in Boolean Formulas}

\author{Alexander Mozeika$^1$}
\author{David Saad$^1$}
\author{Jack Raymond$^2$}

\affiliation{ $^1$The Neural Computing Research Group, Aston University,
Birmingham B4 7ET, UK.
\\ $^2$Department of Physics, The Hong Kong University
of Science and Technology, Clear Water Bay, Hong Kong, China.}

\date{\today}

\begin{abstract}
Computing circuits composed of noisy logical gates and their ability to represent arbitrary Boolean functions with a given level of error are investigated within a statistical mechanics setting. Bounds on their performance, derived in the information theory literature for specific gates, are straightforwardly retrieved, generalized and identified as the corresponding typical-case phase transitions. This framework paves the way for obtaining new results on error-rates, function-depth and sensitivity, and their dependence on the gate-type and noise model used.
\end{abstract}
\maketitle
Noise is inherent in most forms of computing and its impact is more
dramatic as the computing circuits become more complex and of
large scale~\cite{Borkar}. Classical computing circuits
based on electromagnetic components suffer from thermal noise and
production errors, quantum computers suffer from decoherence,
whilst an understanding of noisy processes, inherent in neural
networks and biological systems, remains poorly understood.

The first model of noisy computation was proposed by von Neumann~\cite{VonNeumann} who used Boolean circuits composed of $\epsilon$-noisy gates to gain insight into the robustness of biological neuronal networks. A \emph{circuit} in this context is a directed acyclic graph in which nodes of in-degree zero are either Boolean constants or references to arguments, nodes of in-degree $k\!\geq\!1$ are gates computing Boolean functions of $k$ arguments and nodes of out-degree
zero represent circuit outputs. A \emph{formula} is a single-output circuit
in which the output of each gate is input to at most one
gate. An $\epsilon$-noisy gate computes a Boolean function $\alpha:\{-1,1\}^k\!\rightarrow\!\{-1,1\}$, but for each input $\vecSpin\!\in\!\{-1,1\}^k$ there is an error probability $\epsilon$ such that $\alpha(\vecSpin)\!\rightarrow\!-\!\alpha(\vecSpin)$; we consider the error probability to be independent for each gate. A noisy circuit with $\epsilon\!>\!0$ represents a given deterministic function with a maximum error probability $\delta$ over all possible circuit inputs determining its reliability. Von Neumann showed that reliable computation,  with $\delta\!<\!1/2$, is possible~\cite{VonNeumann} for small $\epsilon$ values and specific gates, and demonstrated how reliability can be improved using $\epsilon$-noisy gates only.
In a more recent analysis Pippenger~\cite{Pippenger:RC} demonstrated that formulae only compute reliably up to a certain threshold in the gate error rate, and that
reliable computation with noisy elements requires strictly greater depth.
These bounds have subsequently been refined~\cite{Feder:RC,Hajek:MTN,Evans:MTNK}, and developed to include
circuits~\cite{Feder:RC}.

Random Boolean functions play an important role in information theory as they allow for the exploration of average case properties~\cite{Brodsky}, in contrast to the traditionally-studied worst-case scenario. The generation of typical functions, sampled uniformly over the space of Boolean functions, is a research area in its own right as most conventional methods focus on the \emph{ability} to construct arbitrary functions using basic gates or procedures, but typically result in highly uncharacteristic functions when generated at random~\cite{Lefmann,Chauvin,Gardy}.
Here we use a growth process where one defines an initial distribution over a set of simple Boolean formulae; these are then combined repeatedly by Boolean connectives to define new formulae. One such process~\cite{Savicky} uses only a single Boolean connective to show that, under very broad conditions, the probability of random functions computed by formulae of depth $\ell$ tends to the uniform distribution over all $n$-variable Boolean functions as $\ell\!\rightarrow\!\infty$~\cite{Savicky}.

In this Letter we show how models of random formulae can be mapped onto a physical framework and employ methods of statistical physics, developed specifically to analyze the \emph{typical} behavior of random disordered systems, to gain insight into the behavior of noisy Boolean random formulae. The stability of the circuit towards input-layer perturbations and its dependence on the input magnetization are studied to establish the main characteristics of the generated formulae. To investigate the properties of noisy circuits we consider two copies of the same topology with different temperatures ($1/\beta$), representing the noisy ($\beta\!<\!\infty$) and noiseless ($\betahat\!\rightarrow\!\infty$) versions of the same circuit. We show that the typical-case macroscopic behavior observed corresponds straightforwardly to the bounds obtained in the information theory literature for specific cases~\cite{VonNeumann}-\cite{Evans:MTNK}. Being very general, the framework is extended to consider further properties of random Boolean formulae for different gates and their dependence on error level and formula depth.
\begin{figure}[t]
\vspace*{-11mm} \hspace*{-0mm} \setlength{\unitlength}{0.25mm}
\begin{picture}(350,210)
\put(0,0){\includegraphics[height=160\unitlength,width=280\unitlength]{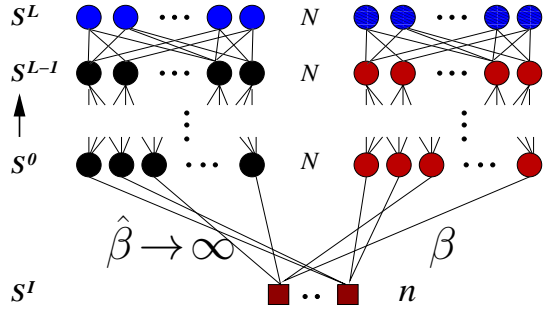}}
\put(50,25){\Large{$\hat{\beta}\!\rightarrow\!\infty$}}
\put(220,25){\Large{$\beta$}}
\end{picture}
 \vspace*{0mm}
\caption{(Color online) The model of two coupled systems with identical topology and different inverse temperatures $\beta$ and $\hat{\beta}\!\rightarrow\!\infty$. Gates are indicated by circles and $S^I$ by squares.\label{fig:0} \vspace*{-0.3cm}
}
\end{figure}

The noisy computation model considered here, shown in Fig.~\ref{fig:0}, is a feed-forward layered $N\!\times\! (L\!+\!1)$ Boolean circuit. The layers in the circuit are numbered from $0$ (input) to $L$ (output). Each layer $\ell\in\{1,..,L\}$ in the circuit is composed of exactly $N$ $\epsilon$-noisy, $k$-ary Boolean gates. Due to gate-noise, the $i$-th gate in the $\ell$-th layer operates in a stochastic manner according to the microscopic law
\begin{equation}
\Prob(\Spin_{i}^\ell\vert \Spin_{i_1}^{\ell-1},\ldots,\Spin_{i_k}^{\ell-1})\!=\!\frac{\rme^{\beta \Spin_{i}^\ell\alpha(\Spin_{i_1}^{\ell-1},\ldots,\Spin_{i_k}^{\ell-1})}}{2\cosh[\beta\alpha(\Spin_{i_1}^{\ell-1},\ldots,\Spin_{i_k}^{\ell-1})]}\label{eq:micro}
\end{equation}
where $\beta$ relates to the gate noise $\epsilon$ via $\tanh\beta\!=\!1\!-\!2\epsilon$. The gate-output $\Spin_{i}^\ell$ is completely random/deterministic when $\beta\!\rightarrow\!0/\infty$, respectively. The model is acyclic by definition so that given the state of the layer $\ell$ the gates of layer $\ell\!+\!1$ operate independently of each other. This suggests that the probability of the microscopic state $\vecSpin^0,..,\vecSpin^L$, where $\vecSpin^\ell\!\in\!\{-1,1\}^N$, is a product of (\ref{eq:micro}) over circuit sites and layers. The joint probability of microscopic states in two systems of identical topology but different gate-noise is
\begin{equation}
\Prob[\{\vecSpin^\ell\};\{\vecSpinhat^\ell\}]\!=\!\Prob(\vecSpin^0,\vecSpinhat^0\vert \vecSpin^I)\prod_{\ell=1}^L \Prob(\vecSpin^\ell\vert\vecSpin^{\ell\!-\!1})P(\vecSpinhat^\ell\vert\vecSpinhat^{\ell\!-\!1})\label{eq:PathProb}
\end{equation}
where
\begin{equation}
\Prob(\vecSpin^\ell\vert\vecSpin^{\ell\!-\!1})\!=\!\prod_{i=1}^N\frac{\rme^{\beta \Spin_{i}^\ell\sum_{j_1,..,j_k}^N A_{j_1,..,j_k}^{\ell,i}\alpha(\Spin_{j_1}^{\ell\!-\!1}\!\!,..,\Spin_{j_k}^{\ell\!-\!1})}}{2\cosh[\beta\sum_{j_1,..,j_k}^N \!A_{j_1,..,j_k}^{\ell,i}\alpha(\Spin_{j_1}^{\ell\!-\!1}\!\!,..,\Spin_{j_k}^{\ell\!-\!1})]}.\label{eq:layerProb}
\end{equation}
The adjacency tensor $A_{j_1,..,j_k}^{\ell,i}\!=\!1$ when it encodes a connection from outputs $j_1,..,j_k$ in layer $\ell\!-\!1$ to gate-input $i$ in layer $\ell$, and 0 otherwise; with $\{i\!=\!1,..,N;\ell\!=\!1,..,L\}$. The conditional probability $\Prob(\vecSpinhat^\ell\vert\vecSpinhat^{\ell\!-\!1})$ is the same as (\ref{eq:layerProb}) but with $\beta\!\rightarrow\!\betahat$. The source of disorder in our model are the random connections and boundary conditions. Random connections are generated by selecting the $i$-th gate at layer $\ell$ and sampling exactly $k$ indices, which point to outputs of layer $\ell\!-\!1$, uniformly from the set of all possible (unordered) indices $\{i_1,..,i_k\}$. This is carried out repeatedly and independently for all gates and layers giving rise to the adjacency tensor probability  $\Prob(A_{j_1,..,j_k}^{\ell,i})\!=\!\frac{1}{N^k}\delta_{A_{j_1,..,j_k}^{\ell,i};1}\!+\!(1\!-\!\frac{1}{N^k})\delta_{A_{j_1,..,j_k}^{\ell,i};0}$. To cater for a possible higher level of correlation, the $0$-layer boundary conditions are generated by selecting randomly members of the finite set $S^I\!=\!\{\Spin^{I}_1,..,\Spin^{I}_n\}$; the indices $x_{i}$ are sampled uniformly with $P(x_{i})\!=\!1/n $ and assigned to the input layer. This leads to the random boundary conditions
$
\Prob(\vecSpin^0,\vecSpinhat^0\vert \vecSpin^I)=\prod_{i=1}^N \delta_{S^0_{i};S^I_{x_{i}}}\delta_{\hat{S}^0_{i};S^0_{i}}~. 
$

The structure of the probability distribution (\ref{eq:PathProb}) is similar to the evolution of disordered Ising spin systems~\cite{ParDyn} if layers are regarded as discrete time-steps of parallel dynamics. The generating functional method~\cite{dD} provides
\begin{eqnarray}
Z[\vecPsi;\vecPsihat]&=&\left\langle\rme^{-\rmi\sum_{\ell,i}\{\psi_i^{\ell} S_{i}^{\ell}+\Psihat_i^{\ell} \hat{S}_{i}^{\ell}\}}\right\rangle~,\label{eq:GF}
\end{eqnarray}	
where $\langle\ldots\rangle$ denotes the average generated by (\ref{eq:PathProb}). The generating functional (\ref{eq:GF}), regarded also as a characteristic function, is used to compute moments of (\ref{eq:PathProb}) by taking partial derivatives with respect to the generating fields $\{\psi_i^{\ell},\Psihat_j^{\ell^\prime}\}$, e.g. $\langle S_i^{\ell}\hat{S}_j^{\ell^\prime}\rangle=-\lim_{\vecPsi,\vecPsihat\rightarrow\nullvec} \frac{\partial^2}{\partial_{\psi_i^\ell}\partial_{\hat{\psi}_j^{\ell^\prime}}}Z[\vecPsi;\vecPsihat]$.  To compute $Z[\vecPsi;\vecPsihat]$ we assume that for $N\rightarrow\infty$ the system becomes self-averaging, i.e. $Z=\overline{Z}$, where $\overline{\cdots}$ is the disorder average. Furthermore,  the normalization property $Z[\nullvec;\nullvec]=1$ allows one to average over the disorder $\overline{Z}$ directly, giving rise to the macroscopic observables
	\begin{eqnarray}
m(\ell)\!=\!\frac{1}{N}\sum_{i=1}^N\overline{\langle S_i^\ell\rangle},~~~~~~~~
C(\ell)=\frac{1}{N}\sum_{i=1}^N\overline{\langle S_i^\ell\hat{S}_i^{\ell}\rangle}~,\label{def:observ}
\end{eqnarray}
the average layer activity (magnetization) $m(\ell)$ on layer $\ell$ and overlap $C(\ell)$ between the two systems. Averaging (\ref{eq:GF}) over the disorder leads to the saddle-point integral
$
\overline{Z[\ldots]}\!=\!\int\{\rmd \vecProb \rmd\vecProbhat \rmd \vecOmega \rmd\vecOmegahat\}\rme^{N\Psi[\vecProb,\vecProbhat; \vecOmega ,\vecOmegahat]}
$
where $\Psi$ is
\begin{eqnarray}
\Psi&=&\rmi\sum_\ell\sum_{ S,\hat{S}}\Probhat^\ell( S,\hat{S})\Prob^\ell( S,\hat{S})\nonumber \\&+&\sum_m\Prob(m)\log \sum_{\{S^\ell,\hat{S}^\ell\}}\M_m[\{S^\ell,\hat{S}^\ell\}] \label{eq:saddle}
\end{eqnarray}
and $\M_m$ is an effective single-site measure (after removing the fields $\vecPsi,\vecPsihat$)
\begin{widetext}
\begin{eqnarray}
M_m[ \{S^\ell,\hat{S}^\ell\}]&=&\delta_{S^0; S^I_m}
\delta_{\hat{S}^0;S^0}\prod_{\ell=0}^{L-1}\Bigg\{\sum_{\{S_j,\hat{S}_j\}}\prod_{j=1}^{k}\left[\Prob^\ell( S_j,\hat{S}_j)\right]\rme^{-\rmi\hat{P}^\ell(S^\ell,\hat{S}^\ell)}\nonumber\\
&&\times
\frac{\rme^{\beta  S^{\ell+1}\alpha(S_1,..,S_k)}}{2\cosh[\beta\alpha(S_1,..,S_k)]}\frac{\rme^{\hat{\beta} \hat{S}^{\ell+1}\alpha(\hat{S}_1,..,\hat{S}_k)}}{2\cosh[\hat{\beta}\alpha(\hat{S}_1,..,\hat{S}_k)]}\Bigg\}.\label{eq:M}
\end{eqnarray}
\end{widetext}
For $N\!\rightarrow\!\infty$ the
averaged generating functional is dominated by the extremum of $\Psi$. Functional variation with respect to the order parameter $\hat{P}^\ell(S^\ell,\hat{S}^\ell)$ provides the saddle-point equation $
%
\Prob^\ell( S,\hat{S})\!=\!\sum_m\Prob(m)\left\langle\delta_{ S^\ell; S}\delta_{\hat{S}^\ell;\hat{S}}\right\rangle_{M_m},
%
$
where $\langle\cdots\rangle_{M_m}$ is the average with respect to (\ref{eq:M}). The physical meaning of $\Prob^\ell( S,\hat{S})$ relates to the averaged joint probability of nodes in the two systems $\Prob^\ell( S,\hat{S})=\lim_{N\rightarrow\infty}\frac{1}{N}\sum_{i=1}^N\overline{\langle\delta_{S_i^\ell;S}\delta_{\hat{S}_i^\ell;\hat S}\rangle\vert_{S^I}}$, while the conjugate order parameter, which ensures normalization of $\Prob^\ell( S,\hat{S})$, vanishes. This simplifies our effective measure (\ref{eq:M}) for computing the macroscopic observables, yielding
\begin{eqnarray}
m(\ell\!+\!1)\!&=&\!\sum_{\{S_j\}}\prod_{j=1}^{k}\left[\frac{1}{2}\{1\!+\! S_jm(\ell)\}\right]\tanh[\beta \alpha(S_1,..,S_k)]\nonumber \label{eq:m}\\
C(\ell\!+\!1)\!&=&\!\!\!\!\!\sum_{\{S_j,\hat{S}_j\}}\!\prod_{j=1}^{k}\biggl[\frac{1}{2}\bigl\{1\!+\! S_j m(\ell)\!+\!\hat{S}_j \hat m(\ell) \!+\! \!S_j\hat  S_j C(\ell)\bigr\}\biggr]\nonumber\\ \!&\times&\!
\tanh[\beta \alpha(S_1,..,S_k)]\tanh[\hat\beta \alpha(\hat{S}_1,..,\hat{S}_k)].\label{eq:Cl}
\end{eqnarray}
The magnetization $\hat{m}(\ell)$ is computed by (\ref{eq:m}) using $\hat{\beta}$; initial conditions are $m(0)\!=\!\hat{m}(0)\!=\!\frac{1}{\vert S^I\vert}\sum_{S\in S^I} \! \Spin,~C(0)\!=\!1$.

The connectivity profile considered here results in a simple set of equations. The macroscopic behavior of the two systems is completely determined by the set of observables $\{m(\ell),\hat{m}(\ell),C(\ell)\}$ through the order parameter $\Prob^\ell( S,\hat{S})\!=\!\frac{1}{2}(1\!+\! S m(\ell)\!+\!\hat{S}\hat m(\ell)\!+\! S\hat{S} C(\ell))$, while the single system behavior is dominated by $\{m(\ell)\}$. Furthermore, since $\langle\prod_m\Spin^\ell_{i_m}\rangle\!\rightarrow\!\prod_m\langle\Spin^\ell_{i_m}\rangle$ for finite $m$, the spins in layer $\ell$ are uncorrelated when $N\!\rightarrow\!\infty$; this is due to the fact that the $i$-th site is a root of a full $k$-ary tree, which grows from the input layer and points to Boolean variables in the set $S^I$. Loops in the circuit are rare, so that trees can be regarded as random independent Boolean formulae for a given input. The output of a typical formula at layer $\ell$ is determined by  $\Prob^\ell(S)$.

The order parameter $C(\ell)$ and the normalized Hamming distance $D(\ell)$  between  states $\vecSpin^\ell$ and $\vecSpinhat^\ell$ are related via the identity $D(\ell)\!=\!\frac{1}{2}(1\!-\!C(\ell))$. This gives rise to the measure $\Delta(\ell)\!=\!\lim_{\beta,\hat{\beta}\!\rightarrow\!\infty}D(\ell)$, for the circuit's sensitivity with respect to its input. The probability
$\Prob(S^\ell_i\!\neq\!\hat{S^\ell}_i)$ for any node, which relates to the Hamming distance $D(\ell)$, facilitates the estimate of the noisy circuit's $\ell$-layer error probability $\delta(\ell)\!=\!\max_{S^I}\lim_{\hat{\beta}\!\rightarrow\!\infty}D(\ell)$, comparing the noisy and noiseless node values for all inputs.  Obviously, in the absence of noise $\delta(\ell)\!=\!0,~\forall\ell$.

\begin{figure}[t]
\vspace*{0mm} \hspace*{-0mm} \setlength{\unitlength}{0.27mm}
\begin{picture}(350,210)
\put(157,110){\includegraphics[height=100\unitlength,width=160\unitlength]{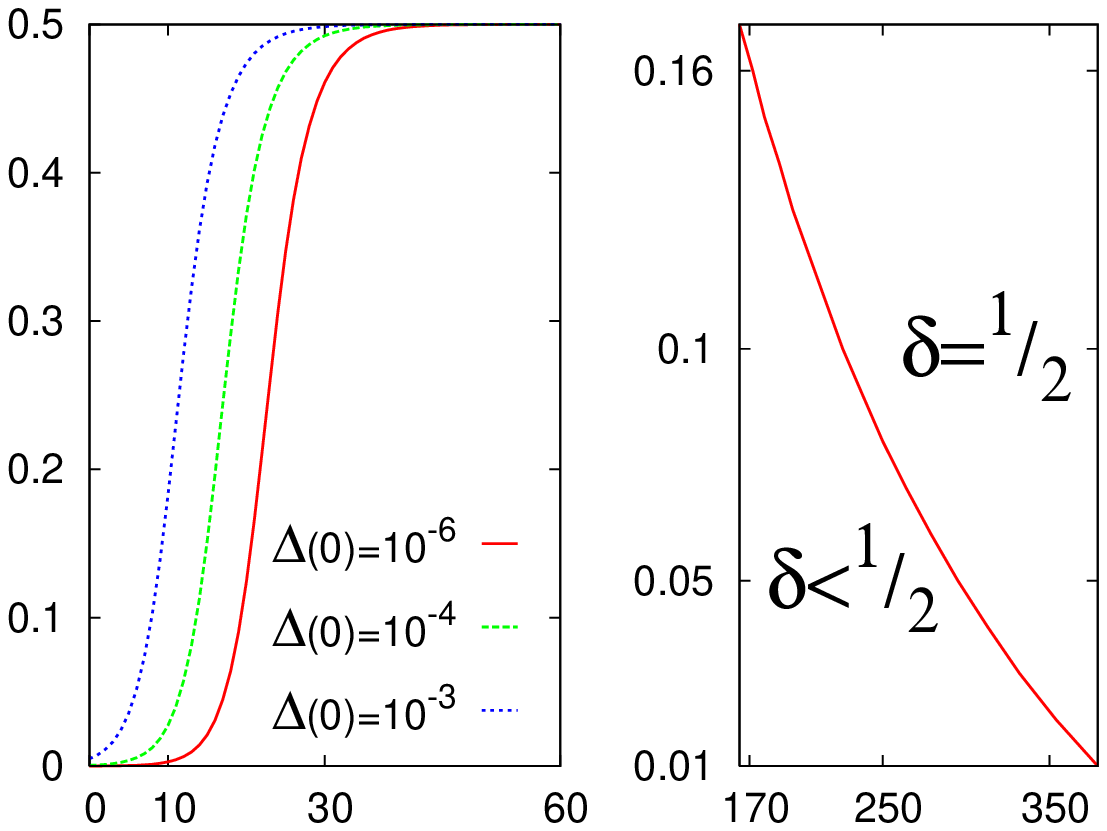}}
\put(0,110){\includegraphics[height=100\unitlength,width=160\unitlength]{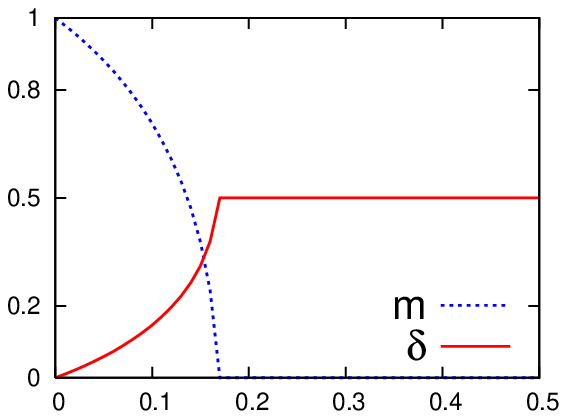}}
\put(159,0){\includegraphics[height=100\unitlength,width=160\unitlength]{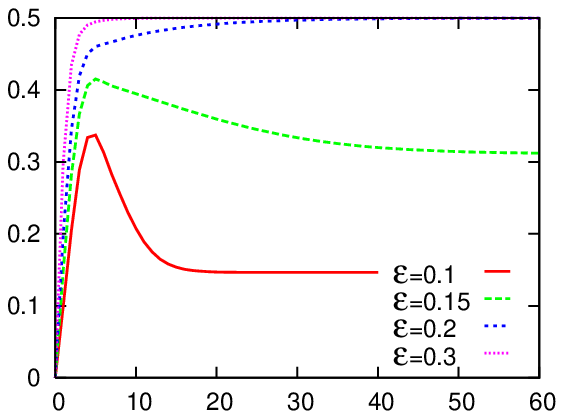}}
\put(245,-7){$\ell$}
\put(0,0){\includegraphics[height=100\unitlength,width=160\unitlength]{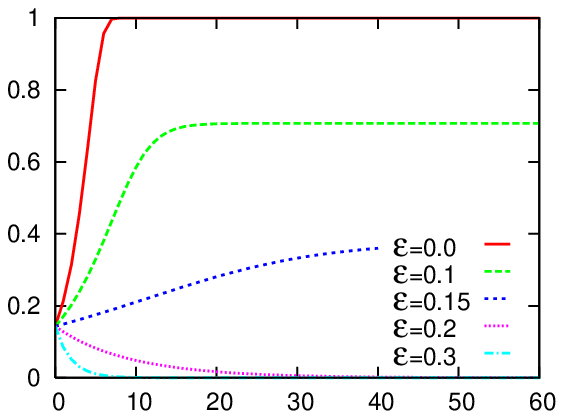}}
\put(-3,65){$m$}
\put(85,-7){$\ell$} \put(160,175){$\Delta$}\put(165,65){$\delta$}
\put(250,178){$\epsilon$}
\put(85,103){$\epsilon$}\put(208,101){$\ell$}\put(280,101){\scriptsize{$L$}} 
\put(130,190){$(a)$}\put(221,190){$(b)$}\put(290,190){$(c)$}\put(130,80){$(d)$}\put(290,80){$(e)$}
\end{picture}
 \vspace*{0mm}
\caption{(Color online) Properties of the MAJ-3 gate: (a) Magnetization $m$ and output error $\delta$ as a function  of gate noise $\epsilon$. (b) Sensitivity of $\Delta(\ell)$ to input mismatch $\Delta(0)$ for $m(0)\!=\!0$. (c) Phase diagram for gate noise $\epsilon$ at layer $L$. For the MAJ-7 \emph{function}, evolution of: (d) Magnetization $m$. (e) Output error $\delta$. \label{fig:1} \vspace*{-0.5cm}
}
\end{figure}

To obtain results for a specific case, which could be compared against those obtained in the information theory literature, we apply equations~(\ref{eq:Cl}) for a particular Boolean gate $\alpha$, the $k$-input majority gate  (MAJ-$k$). The reasons for choosing this gate are twofold. Firstly, it was proved~\cite{Hajek:MTN,Evans:MTNK} to be optimal for noisy computation in formulae. Secondly, a formula constructed at random using majority gates can in principle compute any Boolean function~\cite{Savicky} with \emph{uniform probability}. A convenient representation of the MAJ-$k$ gate is of the form $\text{MAJ}(S_1,..,S_k)\!=\!\sgn[\sum_{j=1}^k S_j]$ with odd $k$. For the particularly simple example MAJ-3 one obtains for $\hat{\beta}\!\rightarrow\!\infty$
\begin{eqnarray}
m(\ell\!+\!1)\!&=&\!\frac{1}{2}\tanh\beta\,[3m(\ell)\!-\!m^3(\ell)]\label{eq:m3}\\
C(\ell\!+\!1)\!&=&\!\tanh\beta\,\Big[\frac{3}{2}m(\ell)\hat m(\ell)\!-\!\frac{3}{4}C(\ell)m^2(\ell) \nonumber\\ \!&-&\!\frac{3}{4}C(\ell)\hat m^2(\ell)
\!+\!\frac{3}{4}C(\ell)\!+\!\frac{1}{4}C^3(\ell)\Big]~.\label{eq:C3}
\end{eqnarray}
Insight on the functions implemented and the gate noise threshold can be obtained from equation~(\ref{eq:m3}), which describes the evolution of the magnetization from layer to layer. When expanded around the stationary solution $m(\infty)\!=\!0$ it identifies the critical noise value $\epsilon^*\!=\!1/6$, identical to the results of~\cite{VonNeumann,Hajek:MTN}, below which the (unordered) $m(\infty)\!=\!0$ solution becomes unstable and two stable (ordered) solutions $m(\infty)\!=\!\pm\!\sqrt{\frac{1\!-\!6\epsilon}{1\!-\!2\epsilon}}$ emerge.
Studying the joint dynamics of (\ref{eq:m3}-\ref{eq:C3}) shows that for
$\epsilon\!>\!1/6$ the magnetization decays to $0$ (exponentially) while for  $\epsilon\!<\!1/6$ the stationary solutions appear, corresponding to the positive and negative initial magnetizations $m(0)$, respectively. The boundary separating these phases, shown in Fig.~\ref{fig:1}a, identifies the noise-level below which the circuit can preserve one bit of input information $S^I\!=\!\{S\}$ for arbitrarily many layers; the error probability  $\Prob^\ell(\!-\!\Spin)\!=\!\frac{1}{2}(1\!-\!\Spin m(\ell))$ measures how well it is preserved after $\ell$  layers. Less complicated functions (fewer layers) can be computed with higher gate noise.

The analysis can easily accommodate other gates, in particular MAJ-$k$. Using similar arguments one identifies the critical noise level $\epsilon^*\!=\! 1/2\!-\!2^{k\!-\!2}/\binom{k\!-\!1}{(k\!-\!1)/2}$ below which two stable solutions emerge. Computing formulae with limited error $\delta$ above the critical noise level $\epsilon^*$, identical to the threshold reported in~\cite{Evans:MTNK}, becomes infeasible. Similarly, the noise threshold for formulae constructed of NAND gates identifies a threshold noise level $\epsilon^*\!=\!(3\!-\!\sqrt{7})/4$, identical to the one derived in~\cite{Evans:MTN}.

General properties of average formulae can be straightforwardly obtained from the site probability of average formulae $\Prob^\ell(\Spin)$ at layer $\ell$.
Stationary solutions in the noiseless case show $m(\infty)\!=\!\pm\!1$, in correspondence  to the sign of the initial magnetization; giving rise to biased function outputs. For $m(0)\!=\!0$ one obtains $m(\infty)\!=\!0$, so that each site of the model can be associated with some random Boolean function output, evaluating to $\pm\!1$ with equal probability. Consequently, depending on the initial conditions, formulae converge to a \emph{single} Boolean function or to the uniform distribution over some set of functions~\cite{Brodsky}. Our result is consistent with majority gate growth process~\cite{Savicky,Brodsky} where for input $S^I\!=\!\{\!-\!1,1,S_1^I,..,S_n^I,\!-\!S_1^I,..,\!-\!S_n^I\}$ stationary state formulae compute all Boolean functions of $n$ variables while for $S^I\!=\!\{\!-\!1,1,S_1^I,..,S_n^I\}$ (also without $\!-\!1,1$) they converge to the MAJ-$n$ function (odd $n$) or to the uniform distribution over slice functions (even $n$)~\cite{Brodsky}. Convergence to the stationary solution $m(\infty)$ is at depth $O(\log n)$ for $m(0)\!=\!1/n$ where $n\!\in\!\mathbb{N}$ in agreement with~\cite{Brodsky}.

\emph{Function} error-rates can be calculated through the study of equation~(\ref{eq:C3}) describing the evolution of the overlap between the two systems. Initial conditions are the same for both systems $m(0)\!=\!\hat m(0)$ and $C(0)\!=\!1$. The magnetization in the noisy system ($\epsilon\!\leq\!1/6$) converges to $m(\infty)\!=\!\pm\!\sqrt{\frac{1\!-\!6\epsilon}{1\!-\!2\epsilon}}$, depending on the sign of $m(0)$. Using these stationary values and equation~(\ref{eq:C3}) we find $C(\infty) \left( 7\!-\!18\,\epsilon \right) \!-\! \left( 1\!-\!2\,\epsilon \right) {C^{3}(\infty)}
\!=\!\pm 6 \sqrt { \left( 1\!-\!2\,\epsilon \right)  \left( 1\!-\!6\,\epsilon
\right) }$ leading to the error probability $\delta(\infty)$ plotted in Fig.~\ref{fig:1}a.

The stationary solution  $C(\infty)\!=\!1$ of equation~(\ref{eq:C3}) for initial conditions $m(0)\!=\!0$, $C(0)\!=\!1$ and $\epsilon\!=\!0$ is unstable under perturbations to $C(0)$, resulting in the stable stationary state $C(\infty)\!=\!0$. Consequently, the circuit is input-sensitive leading to an increasing Hamming distance $\Delta(\ell)$ for small perturbations $\Delta(0)$ as shown in Fig.~\ref{fig:1}b. For $\epsilon\!>\!0$ the circuit amplifies the noise and $\delta(L)$ grows but remains limited for sufficiently small $\epsilon$ as shown in Fig.~\ref{fig:1}c.

%
%

To examine the computation performed at layer $\ell$ we consider the input set $S^I\!=\!\{S_1,..,S_7\}$, corresponding to the function MAJ-7 for the noiseless case, with lowest possible initial magnetization $m(0)\!=\!1/7$ where changes between layers are smallest.
Figure~\ref{fig:1}d shows the magnetization $m(\ell)$ for different gate-noise levels; the convergence-rate decreases with increasing $\epsilon$.
%
%
Close to the critical value the difference equation (\ref{eq:m3}) can be approximated by the differential equation
$\frac{\mathrm{d}}{\mathrm{d} \ell}m(\ell)\!=\!-\!m(\ell)\!+\!\frac{1}{2}(1\!-\!2\epsilon)[3m(\ell) \!-\! m^3(\ell)]$ for continuous $\ell$. Its solution close to the phase boundary, obtained by expanding $\epsilon\!=\!1/6\!+\!\Delta\epsilon$ where $\vert\Delta\epsilon\vert\!\ll\!1$, exhibits exponential convergence $\vert m(\ell)\!-\!m(\infty)\vert\approx\rme^{-\text{const}\Delta\epsilon\ell}$.

The function error $\delta(\ell)$, shown in Fig.~\ref{fig:1}e for different $\epsilon$ values, exhibits two distinct stages in the dynamics. Initially, the error increases until it reaches its maximum value at $\ell\!=\!5$, before the MAJ-7 function is computed exactly at $\ell\!=\!8$ for $\epsilon\!=\!0$ (see Fig.~\ref{fig:1}d); the location of this maximum is independent of $\epsilon$. This suggests that gate-inputs at layers $\ell\!\leq\!5$ are non-uniform, contributing to noise-amplification, but become more uniform later leading to  noise-suppression and decreasing error. As we approach $\epsilon^*$ the number of layers needed for the error to reach stationarity increases; in the region $\epsilon\!=\!1/6\!\pm\!\Delta\epsilon$ it can be estimated from the asymptotic form derived for $m(\ell)$. The dynamic behavior of the error changes to monotonically increasing at $\epsilon^0\!=\!\frac{1}{2}\left[\frac{1\!-\!m^2(0)}{3\!-\!m^2(0)}\right]$ above which noise cannot be reduced by additional layers. For $\epsilon\!\gg\!1/6$ the error evolution becomes strictly monotonic it relaxes to its stationary value $1/2$ exponentially fast.

%
%
By mapping the problem of noisy computation onto a physical framework, we retrieved many of the existing bounds and extended them to include arbitrary gates and/or distribution of gates. In addition, we calculated the level of error and function-bias expected at any depth, the sensitivity to input perturbations and expected convergence rate depending on the input bias, gate properties and gate-noise level. This framework enables one to discover typical properties of noisy computation that are inaccessible via traditional methods of information theory and will undoubtedly contribute to exciting new discoveries. For instance, one can show that systems composed of the biologically-inspired perceptron-like gates are more robust against gate noise than other logical gates and study the effect of hard (systematic) noise.

\begin{acknowledgments} Support by the
Leverhulme trust is acknowledged.
\end{acknowledgments}
%

\begin{thebibliography}{14}
\expandafter\ifx\csname natexlab\endcsname\relax\def\natexlab#1{#1}\fi
\expandafter\ifx\csname bibnamefont\endcsname\relax
  \def\bibnamefont#1{#1}\fi
\expandafter\ifx\csname bibfnamefont\endcsname\relax
  \def\bibfnamefont#1{#1}\fi
\expandafter\ifx\csname citenamefont\endcsname\relax
  \def\citenamefont#1{#1}\fi
\expandafter\ifx\csname url\endcsname\relax
  \def\url#1{\texttt{#1}}\fi
\expandafter\ifx\csname urlprefix\endcsname\relax\def\urlprefix{URL }\fi
\providecommand{\bibinfo}[2]{#2}
\providecommand{\eprint}[2][]{\url{#2}}

\bibitem[{\citenamefont{Borkar}(2005)}]{Borkar}
\bibinfo{author}{\bibfnamefont{S.}~\bibnamefont{Borkar}},
  \bibinfo{journal}{IEEE Micro} \textbf{\bibinfo{volume}{25}},
  \bibinfo{pages}{10} (\bibinfo{year}{2005}).

\bibitem[{\citenamefont{Von~Neumann}(1956)}]{VonNeumann}
\bibinfo{author}{\bibfnamefont{J.}~\bibnamefont{Von~Neumann}},
  \emph{\bibinfo{title}{Probabilistic logics and the synthesis of reliable
  organisms from unreliable components}} (\bibinfo{publisher}{Princeton
  University Press}, \bibinfo{address}{Princeton, NJ}, \bibinfo{year}{1956}),
  p. \bibinfo{pages}{43–98}.

\bibitem[{\citenamefont{Pippenger}(1988)}]{Pippenger:RC}
\bibinfo{author}{\bibfnamefont{N.}~\bibnamefont{Pippenger}},
  \bibinfo{journal}{IEEE Trans. Inf. Theory} \textbf{\bibinfo{volume}{34}},
  \bibinfo{pages}{194} (\bibinfo{year}{1988}).

\bibitem[{\citenamefont{Feder}(1989)}]{Feder:RC}
\bibinfo{author}{\bibfnamefont{T.}~\bibnamefont{Feder}}, \bibinfo{journal}{IEEE
  Trans. Inf. Theory} \textbf{\bibinfo{volume}{35}}, \bibinfo{pages}{569}
  (\bibinfo{year}{1989}).

\bibitem[{\citenamefont{Hajek and Weller}(1991)}]{Hajek:MTN}
\bibinfo{author}{\bibfnamefont{B.}~\bibnamefont{Hajek}} \bibnamefont{and}
  \bibinfo{author}{\bibfnamefont{T.}~\bibnamefont{Weller}},
  \bibinfo{journal}{IEEE Trans. Inf. Theory} \textbf{\bibinfo{volume}{37}},
  \bibinfo{pages}{388} (\bibinfo{year}{1991}).

\bibitem[{\citenamefont{Evans and Schulman}(2003)}]{Evans:MTNK}
\bibinfo{author}{\bibfnamefont{W.}~\bibnamefont{Evans}} \bibnamefont{and}
  \bibinfo{author}{\bibfnamefont{L.}~\bibnamefont{Schulman}},
  \bibinfo{journal}{IEEE Trans. Inf. Theory} \textbf{\bibinfo{volume}{49}},
  \bibinfo{pages}{3094} (\bibinfo{year}{2003}).

\bibitem[{\citenamefont{Brodsky and Pippenger}(2005)}]{Brodsky}
\bibinfo{author}{\bibfnamefont{A.}~\bibnamefont{Brodsky}} \bibnamefont{and}
  \bibinfo{author}{\bibfnamefont{N.}~\bibnamefont{Pippenger}},
  \bibinfo{journal}{Random Struct. Algor.} \textbf{\bibinfo{volume}{27}},
  \bibinfo{pages}{490} (\bibinfo{year}{2005}).

\bibitem[{\citenamefont{Lefmann and Savick\'{y}}(1997)}]{Lefmann}
\bibinfo{author}{\bibfnamefont{H.}~\bibnamefont{Lefmann}} \bibnamefont{and}
  \bibinfo{author}{\bibfnamefont{P.}~\bibnamefont{Savick\'{y}}},
  \bibinfo{journal}{Random Struct. Algor.} \textbf{\bibinfo{volume}{10}},
  \bibinfo{pages}{337} (\bibinfo{year}{1997}).

\bibitem[{\citenamefont{Chauvin et~al.}(2004)\citenamefont{Chauvin, Flajolet,
  Gardy, and Gittenberger}}]{Chauvin}
\bibinfo{author}{\bibfnamefont{B.}~\bibnamefont{Chauvin}},
  \bibinfo{author}{\bibfnamefont{P.}~\bibnamefont{Flajolet}},
  \bibinfo{author}{\bibfnamefont{D.}~\bibnamefont{Gardy}}, \bibnamefont{and}
  \bibinfo{author}{\bibfnamefont{B.}~\bibnamefont{Gittenberger}},
  \bibinfo{journal}{Comb. Probab. Comput.} \textbf{\bibinfo{volume}{13}},
  \bibinfo{pages}{475} (\bibinfo{year}{2004}).

\bibitem[{\citenamefont{Gardy and Woods}(2005)}]{Gardy}
\bibinfo{author}{\bibfnamefont{D.}~\bibnamefont{Gardy}} \bibnamefont{and}
  \bibinfo{author}{\bibfnamefont{A.}~\bibnamefont{Woods}}, in
  \emph{\bibinfo{booktitle}{2005 International Conference on Analysis of
  Algorithms}}, edited by
  \bibinfo{editor}{\bibfnamefont{C.}~\bibnamefont{Mart\'{\i}nez}}
  (\bibinfo{year}{2005}), vol.~\bibinfo{volume}{AD} of
  \emph{\bibinfo{series}{DMTCS Proceedings}}, pp. \bibinfo{pages}{139--146}.

\bibitem[{\citenamefont{Savick\'{y}}(1990)}]{Savicky}
\bibinfo{author}{\bibfnamefont{P.}~\bibnamefont{Savick\'{y}}},
  \bibinfo{journal}{Discrete Math.} \textbf{\bibinfo{volume}{83}}
  (\bibinfo{year}{1990}).

\bibitem[{\citenamefont{Hatchett et~al.}(2004)\citenamefont{Hatchett,
  Wemmenhove, Castillo, Nikoletopoulos, Skantzos, and Coolen}}]{ParDyn}
\bibinfo{author}{\bibfnamefont{J.~P.~L.} \bibnamefont{Hatchett}},
  \bibinfo{author}{\bibfnamefont{B.}~\bibnamefont{Wemmenhove}},
  \bibinfo{author}{\bibfnamefont{I.~P.} \bibnamefont{Castillo}},
  \bibinfo{author}{\bibfnamefont{T.}~\bibnamefont{Nikoletopoulos}},
  \bibinfo{author}{\bibfnamefont{N.~S.} \bibnamefont{Skantzos}},
  \bibnamefont{and} \bibinfo{author}{\bibfnamefont{A.~C.~C.}
  \bibnamefont{Coolen}}, \bibinfo{journal}{J. Phys. A: Math. Gen.}
  \textbf{\bibinfo{volume}{37}}, \bibinfo{pages}{6201} (\bibinfo{year}{2004}).

\bibitem[{\citenamefont{de~Dominics}(1978)}]{dD}
\bibinfo{author}{\bibfnamefont{C.}~\bibnamefont{de~Dominics}},
  \bibinfo{journal}{Phys. Rev. B.} \textbf{\bibinfo{volume}{18}},
  \bibinfo{pages}{4913} (\bibinfo{year}{1978}).

\bibitem[{\citenamefont{Evans and Pippenger}(1998)}]{Evans:MTN}
\bibinfo{author}{\bibfnamefont{W.}~\bibnamefont{Evans}} \bibnamefont{and}
  \bibinfo{author}{\bibfnamefont{N.}~\bibnamefont{Pippenger}},
  \bibinfo{journal}{IEEE Trans. Inf. Theory} \textbf{\bibinfo{volume}{44}},
  \bibinfo{pages}{1299} (\bibinfo{year}{1998}).

\end{thebibliography}

\end{document}